\documentclass[prodmode,acmtopc,leqno]{acmsmall} 

\usepackage[ruled]{algorithm2e}

\SetAlFnt{\small}
\SetAlCapFnt{\small}
\SetAlCapNameFnt{\small}
\SetAlCapHSkip{0pt}
\IncMargin{-\parindent}

\acmVolume{0}
\acmNumber{0}
\acmArticle{}
\acmYear{0000}
\acmMonth{0}

\setcopyright{rightsretained}

\doi{\dots}

\issn{\dots}

\usepackage[usenames,dvipsnames,svgnames,table]{xcolor}
\usepackage{pgfplots}
\usepackage{amsmath,dsfont}
\usepackage[authoryear,square]{natbib}
\usepackage{mathtools}
\DeclareMathOperator{\spn}{span}

\pgfplotsset{compat=newest}

\usepackage{bm}
\usepackage{tikz}
\usepackage[11pt]{moresize}
\usepackage{listings}
\usepackage{caption}
\usepackage{arydshln}
\usepackage[defaultlines=2,all]{nowidow}

\renewcommand{\thefigure}{\arabic{figure}}
\renewcommand{\thetable}{\arabic{table}}

\newcommand\Tstrut{\rule{0pt}{2.2ex}}         
\newcommand\Bstrut{\rule[-1ex]{0pt}{0pt}}   

\usepgfplotslibrary{fillbetween}

\usetikzlibrary{arrows,automata,intersections,calc}
\usetikzlibrary{graphs,graphs.standard,quotes,positioning,arrows}
\usetikzlibrary{decorations.pathreplacing,plotmarks,matrix,fit}
\usetikzlibrary{shapes.symbols}

\tikzset{dot/.style={shape=circle,fill=black!80,scale=0.3}}
\tikzset{brightdot/.style={shape=circle,fill=black!15,scale=0.3}}
\tikzset{bigger_dot/.style={shape=circle,fill=red!50,scale=0.5}}
\tikzset{bigger_brighter_dot/.style={shape=circle,fill=red!20,scale=0.5}}
\tikzset{withtext/.style={fill=white}}
\newcommand{\AxisRotator}[1][rotate=0]{%
    \tikz [x=0.25cm,y=0.5cm,line width=.1ex,-stealth,#1] \draw (0,0) arc (-150:150:1 and 1);%
}
\newcommand*\circled[1]{\tikz[baseline=(char.base)]{
            \node[shape=circle,draw,inner sep=1pt,line width=.2ex] (char) {\bfseries\ssmall #1};}}

\definecolor{PseudocodeKeyword}{rgb}{.75, .17, .67}
\definecolor{PseudocodeComment}{rgb}{.13, .52, 0}
\definecolor{Gray}{rgb}{.70, .70, .70}
\definecolor{UzLGreen}{rgb}{0, .36, .39}
\definecolor{LightRed}{rgb}{0.945, 0.255, 0.275}

\lstdefinestyle{pseudocode}{ %
	backgroundcolor=\color{white},   
 	basicstyle=\linespread{1.14}\ttfamily\footnotesize\selectfont, 
 	breakatwhitespace=false,         
	breaklines=true,                 
 	captionpos=b,                    
 	commentstyle=\color{PseudocodeComment}\selectfont,    
 	deletekeywords={...},            
 	escapeinside={@}{@},             
 	extendedchars=true,              
 	frame=false,	                 
 	keepspaces=true,                 
 	keywordstyle=\color{PseudocodeKeyword}\selectfont,       
 	identifierstyle=\color{black}\selectfont,
 	language=C,                		 
 	otherkeywords={function, par, seq},      
 	numbers=left,                    
 	numbersep=5pt,                   
 	numberstyle=\tiny\color{Gray}\selectfont, 
 	rulecolor=\color{Gray},         
 	showspaces=false,                
 	showstringspaces=false,          
 	showtabs=false,                  
 	stepnumber=1,                    
 	stringstyle=\color{PseudocodeString}\selectfont,     
 	tabsize=2,	                  	 
	morecomment=[l]{//},
  	morecomment=[s]{/*}{*/},
 	morestring=[b]",
  	sensitive=false, 
    belowskip=0em,
}

\begin{document}

\markboth{D.-M.\ Lux, C.\ W\"ulker, and G.\ S.\ Chirikjian}{Parallelization of the FFT on SO(3)}

\title{Parallelization of the FFT on SO(3)\footnote{\,Our C++ implementation is available from the corresponding author.}}
\author{DENIS-MICHAEL LUX
\affil{L\"ubeck University}
CHRISTIAN W\"ULKER\footnote{\,Corresponding author (christian.wuelker@jhu.edu)}
\affil{Department of Mechanical Engineering, Johns Hopkins University}
GREGORY S.\ CHIRIKJIAN
\affil{Department of Mechanical Engineering, Johns Hopkins University}}

\begin{abstract}
In this paper, a work-optimal parallelization of Kostelec and Rockmore's well-known fast Fourier transform and its inverse on the three-dimensional rotation group SO(3) is designed, implemented, and tested. To this end, the sequential algorithms are reviewed briefly first. In the subsequent design and implementation of the parallel algorithms, we use the well-known Forster (PCAM) method and the OpenMP standard. The parallelization itself is based on symmetries of the underlying basis functions and a geometric approach in which the resulting index range is transformed in such a way that distinct work packages can be distributed efficiently to the computation nodes. The benefit of the parallel algorithms in practice is demonstrated in a speedup- and efficiency-assessing benchmark test on a system with 64 cores. Here, for the first time, we present positive results for the full transforms for the both accuracy- and memory-critical bandwidth 512. Using all 64 available cores, the speedup for the largest considered bandwidths 128, 256, and 512 amounted to 29.57, 36.86, and 34.36 in the forward, and 24.57, 26.69, and 24.25 in the inverse transform, respectively.
\end{abstract}

%


\acmformat{D.-M.\ Lux, C.\ W\"ulker, and G.\ S.\ Chirikjian, 2018. Parallelization of the fast Fourier transform on SO(3).}

\maketitle

\section{INTRODUCTION}
Computational harmonic analysis on the three-dimensional rotation group SO(3) has many applications in industry and the natural science disciplines. 
In the technique of \emph{fast rotational matching}, for example, one tries to find the rotation under which a given object resembles another object as closely as possible \cite{Kovacs:2002}. The meaning of `resemblance' between two objects is here determined by the specific context. 
In electron microscopy (EM), for instance, it can be the aim to fit an available high-resolution structure from X-ray crystallography or nuclear magnetic resonance (NMR) spectroscopy into the reconstructed three-dimensional EM image \citep{7164051,fitting,KAM198015}. 
On the other hand, computational harmonic analysis on SO(3) can be used in \emph{molecular replacement} to solve the phase problem in X-ray crystallography \citep{fast_rotation_function_3,fast_rotation_function_2,fast_rotation_function}.
Another application of fast rotational matching in the field of molecular biology is \emph{virtual drug screening} \citep{virtual_screening_1}. 
Here, the goal is to maximize the overlap of the electron density functions of two given biomolecules, \textit{e.g.}, by a mass-center-aligned purely rotational search. The concomitant determination of the maximal overlap is used to evaluate the structural resemblance of both molecules. In this way, it is possible to draw conclusions about similarities in the features 
of the molecules. 
Rotating a biomolecule against another also forms an essential part in the related field of \emph{protein-protein docking} (see, \textit{e.g.}, \citep{protein_docking_2,protein_docking_4,protein_docking_3,protein_docking_1}).
The problem of optimal three-dimensional rotation alignment has been addressed in the field of computer graphics \citep{computer_graphics_1}, too, where also rotation-invariant 
shape descriptors are of interest \citep{computer_graphics_2}. Reversely, the problem can be to estimate a rotation, say, from spherical images \citep{rotation_estimation_2,rotation_estimation_1}. 
Last but not least, computational spectral analysis on SO(3) plays an important role in computational harmonic analysis on the three-dimensional Euclidean motion group SE(3) \citep{KYATKIN2000220}.\pagebreak

The fast Fourier transform on SO(3) (hereinafter abbreviated as FSOFT) and its inverse (iFSOFT) introduced by \citet{Kostelec:Rockmore:2008} nowadays constitute an integral part of many techniques based on, including, or related to fast rotational matching (see \cite{Buelow:2011a,Buelow:2011b} for examples in robotics and automation, \cite{Slabaugh:2008} for an example in industrial design, and \citep{Baboud:2011,Makadia2010,Makadia:2007} for examples in computer vision/graphics). However, a parallelization of these computationally demanding algorithms has not yet been undertaken and presented (though it is noteworthy that protein-protein docking has been accelerated using graphic processors \citep{hex,cuda}). In this sense, the aim of this work is to design and implement such parallelization of the FSOFT and iFSOFT, and to demonstrate the practical applicability of the resulting parallel algorithms.

The remainder of this paper is organized as follows: 
Firstly, in Section \ref{sec:preliminaries}, we briefly review the mathematical theory of the three-dimensional rotation group SO(3) and corresponding discrete Fourier transform. The FSOFT and iFSOFT are then revisited. 
Subsequently, in Section \ref{sec:parallelization}, the parallelization of the sequential algorithms is carried out. Here, we use the Forster (a.k.a.\ PCAM) method \citeyear{forster}, as well as the OpenMP standard (\url{www.openmp.org}). Our parallelization is based on symmetries of the underlying basis functions and a geometric approach in which the resulting index range is transformed so as to allow for an efficient distribution of distinct work packages to the available computation nodes.
Section \ref{sec:benchmark_tests} contains the benchmark test by means of which we demonstrate the practical benefits of our parallelization using our freely available implementation. 
In Section \ref{sec:discussion_and_conclusion}, we discuss the results and give an outlook on future developments.

\section{Preliminaries}\label{sec:preliminaries}
The aim of this section is to familiarize the reader with the sequential fast SO(3) Fourier transform and its inverse of \citet{Kostelec:Rockmore:2008}, in order to prepare for the parallelization in Section \ref{sec:parallelization}. By those already familiar with the FSOFT and iFSOFT, this section can be skipped. In Subsection \ref{subsec:so3}, the three-dimensional rotation group SO(3) is defined. We then introduce the corresponding basis functions, called Wigner-$D$ functions, in Subsection \ref{subsec:wigner}. In Subsection \ref{subsec:dsoft}, the discrete SO(3) Fourier transform and its inverse are described. The fast SO(3) Fourier transform and its inverse are revisited in Subsection \ref{subsec:sofft}.

\subsection{SO(3)}
\label{subsec:so3}

The algebraic group SO(3) of rotations in three-dimensional real space consists of all orthogonal real-valued $3 \times 3$ matrices with determinant one, \textit{i.e.},
\begin{equation*}
\mathrm{SO}(3) \coloneqq \{R \in \mathbb{R}^{3 \times 3} \mid R^\mathrm{T} R = E_3,\: \det R = 1\},
\end{equation*}
where $E_3$ denotes the $3 \times 3$ identity matrix. Each such matrix $R$ is called a \emph{rotation}, because multiplication of a three-dimensional vector (in Cartesian coordinates) by $R$ results in a rotation of this vector around a certain rotation axis. In this context, the group operation (composition) of two rotations is simply matrix multiplication. There are three elementary rotation matrices, $R_x$, $R_y$, and $R_z$, describing rotations about the $x$, $y$, and $z$ axis, respectively. For a given rotation angle $\alpha \in [0, 2\pi)$, they are given by
\begin{equation*}
R_x(\alpha) \coloneqq\!
\begin{bmatrix}
1 & 0 & 0\\
0 & \cos\alpha & -\sin\alpha\\ 
0 & \sin\alpha & \cos\alpha 
\end{bmatrix}\!,~~ 
R_y(\alpha) \coloneqq\!
\begin{bmatrix}
\cos\alpha & 0 & \sin\alpha\\
0 & 1 & 0\\
-\sin\alpha & 0 & \cos\alpha
\end{bmatrix}\!,~~
R_z(\alpha) \coloneqq\!
\begin{bmatrix}
\cos\alpha & -\sin\alpha & 0\\
\sin\alpha & \cos\alpha & 0\\
0 & 0 & 1
\end{bmatrix}\!.
\end{equation*}
In the following, we use the so-called $z$-$y$-$z$ \emph{Euler-angle decomposition}, which is a particular way of parameterizing SO(3). In this decomposition, only two of the above elementary rotation matrices are required, namely $R_y$ and $R_z$:
\emph{
To each rotation $R \in \mathrm{SO}(3)$ we can assign \emph{Euler angles} $\alpha \in [0,2\pi)$, $\beta \in [0,\pi]$, and $\gamma \in [0,2\pi)$, such that}
\begin{equation*}
R = R(\alpha,\beta,\gamma) = R_z(\gamma) R_{y}(\beta) R_{z}(\alpha).
\end{equation*}

\subsection{Wigner-\textit{D} functions}
\label{subsec:wigner}

In this section, we introduce the so-called \emph{Wigner-$D$ functions}, which are special functions mapping from SO(3) to the set of complex numbers. 
These functions can be defined using the Euler angles introduced above 
as
\begin{equation}\label{eq:Wigner_D}
%
D(l,m,m'; \alpha, \beta, \gamma) \coloneqq \exp(- \mathrm{i} m \alpha) \hspace*{1pt} d(l,m,m'; \beta)  \exp(- \mathrm{i} m' \gamma),
\end{equation}
where $l \geq 0$ and $m,m'\! = - \hspace*{2pt} l, \dots, l$, while $d(l,m,m')$ denotes the respective real-valued \emph{Wigner-$d$ function} 
\begin{align*}
d(l, m, m'; \beta) &\coloneqq (-1)^{m+m'} \sqrt{\frac{(l+m')!}{(l+m)!}\frac{(l-m')!}{(l-m)!}}\\ &\,\quad \times \left(\sin\frac{\beta}{2} 
\right)^{m'\!-m}\! \left(\cos\frac{\beta}{2}\right)^{m+m'} P(l-m';m'\!-m,m+m';\cos\beta),
\end{align*} 
where $P(l-m';m'\!-m,m+m')$ denotes the standard Jacobi polynomial of degree $l-m'$ and with exponents $m'\!-m$ and $m+m'$ (cf.\ \cite[Eq.\,22.2.1]{abramowitz_stegun}).
The Wigner-$D$ functions serve as the basis functions for the discrete and fast SO(3) Fourier transform, playing essentially the same role as the monomials $\exp(\mathrm{i} k x)$ in the classical discrete Fourier transform (DFT).

In the following, we consider the linear spaces spanned by the Wigner-$D$ functions with $l$ less than a fixed $B \geq 1$,
\begin{equation*}
H_B \coloneqq \spn \lbrace D(l, m, m') \,:\, l < B;\; m, m'\! = -l,\dots,l \rbrace.
\end{equation*}
These linear combinations of Wigner-$D$ functions are called \emph{bandlimited} functions with \emph{bandwidth} $B$.
Each space $H_B$ is a finite-dimensional subspace of the space $L^2(\mathrm{SO}(3))$ of functions square-integrable over SO(3), inheriting the inner product
\begin{equation*}
\langle f, g \rangle \coloneqq \int_0^{2\pi}\!\! \int_0^{\pi}\! \int_0^{2\pi} f(\alpha, \beta, \gamma) \hspace*{1pt} \overline{g(\alpha, \beta, \gamma)} \hspace{1pt} \mathrm{d}\alpha \hspace{1pt} \sin\beta \hspace{2pt} \mathrm{d}\beta \hspace{2pt} \mathrm{d}\gamma,
\end{equation*}
where $\overline{g}$ denotes the complex conjugate of $g$. It can be shown that the Wigner-$D$ functions satisfy the orthogonality relation \cite[Eq.\,2.7]{Kostelec:Rockmore:2008}
\begin{equation*}
\langle D(j, k , k'), D(l, m, m') \rangle = \frac{8\pi^2}{2l+1} \hspace*{1pt} \delta(j,l) \hspace*{1pt} \delta(k,m) \hspace*{1pt} \delta(k'\!,m').
\end{equation*}
Here, $\delta$ denotes the standard Kronecker-delta symbol.
The Wigner-$D$ functions with $l < B$ thus form an \emph{orthogonal basis} of $H_B$. This property is crucial, because the orthogonality of the Wigner-$D$ functions allows to determine the unique Fourier representation of functions $f \in H_B$ by computing the inner products of $f$ with the basis elements $D(l, m, m')$ for $l < B$ (as explained 
in the upcoming Section \ref{subsec:dsoft}). The most important property of the Wigner-$D$ functions in the context of harmonic analysis on SO(3) is the fact that 
%
%
\emph{the Wigner-$D$ functions constitute an orthogonal basis of $L^2(\mathrm{SO}(3))$}.

To close this section, we collect some useful properties of the Wigner-$d$ functions, which will later be used to evaluate these functions numerically in an efficient manner.
The Jacobi polynomials are orthogonal polynomials. As such, they satisfy a \emph{three-term recurrence relation}. Since the particular Jacobi polynomial $P(l-m';m'\!-m,m+m')$ is included in the Wigner-$d$ function $d(l,m,m')$ as a factor, we find that when $l \geq 1$,
\begin{align}\label{eq:Wigner_d_recurrence}
d(l\!+\!1, m, m'; \beta) 
=\, &\frac{(l\!+\!1)(2l\!+\!1)}{\sqrt{((l\!+\!1)^2-m^2)((l\!+\!1)^2-(m')^2)}}\left(\cos\beta -\frac{mm'}{l(l\!+\!1)}\right)d(l, m, m'; \beta)\\\notag
&- \frac{(l\!+\!1)\sqrt{(l^2-m^2)(l^2-(m')^2)}}{l\sqrt{((l\!+\!1)^2-m^2)((l\!+\!1)^2-(m')^2)}}\hspace*{2pt}d(l\!-\!1, m, m'; \beta).
\end{align}
The following initial cases seed this recursion:
\begin{align*}
d(m, \pm m, m'; \beta)  &= \sqrt{\frac{(2m)!}{(m+m')!(m-m')!}}\left(\cos\frac{\beta}{2}\right)^{\!m \pm m'}\!\left(\pm \sin\frac{\beta}{2}\right)^{\!m \mp m'}\!,\\
d(m'\!, m, \pm m'; \beta)  &= \sqrt{\frac{(2m')!}{(m' + m)!(m'-m)!}}\left(\cos\frac{\beta}{2}\right)^{\!m' \pm m}\!\left(\mp \sin\frac{\beta}{2}\right)^{\!m' \mp m}\!.
\end{align*}
The Wigner-$d$ functions also present some useful symmetries \cite[P.\,59\,f.]{Edmonds:1996}:
\begin{align}\label{eq:Wigner_symmetries}
d(l, m, m'; \beta)
&= (-1)^{m-m'} d(l, -m, -m'; \beta)\\\notag
&= (-1)^{m-m'} d(l, m'\!, m; \beta)\\\notag
&= (-1)^{l-m'} d(l, -m, m'; \pi-\beta)\\\notag
&= (-1)^{l+m}\,\hspace*{1pt} d(l, m, -m'; \pi-\beta)\\\notag
&= (-1)^{l-m'} d(l, -m'\!, m; \pi-\beta)\\\notag
&= (-1)^{l+m}\,\hspace*{1pt} d(l, m'\!, -m; \pi-\beta)\\\notag
&= d(l, -m'\!, -m; \beta)\vphantom{(-1)^{l-m'}}.
\end{align}

\subsection{Discrete SO(3) Fourier transform}
\label{subsec:dsoft}

Let a bandwidth $B \geq 1$ be fixed. Functions $f$ on SO(3) with bandwidth $B$, \textit{i.e.}, functions $f \in H_B$, possess a unique \emph{Fourier representation}
\begin{equation}\label{eq:fourier_representation}
f = \sum_{l=0}^{B-1} \sum_{m,m'=-l}^l f^\circ(l,m,m') \hspace*{1pt} D(l,m,m'),
\end{equation}
where the (generalized) \emph{Fourier coefficients} $f^\circ(l,m,m')$ are given by
\begin{equation*}
f^\circ(l,m,m') \coloneqq \frac{2l+1}{8 \pi^2} \langle f, D(l,m,m') \rangle.
\end{equation*}
The discrete SO(3) Fourier transform and its inverse, and hence also the FSOFT and iFSOFT, are based on the following SO(3) \emph{sampling theorem} \cite[Thm.\,1]{Kostelec:Rockmore:2008}:

\emph{
Let $f \in H_B$. The Fourier coefficients $f^\circ(l,m,m')$ of $f$ obey the quadrature formula
\begin{equation}\label{eq:quadrature}
f^\circ(l,m,m') = \frac{2l+1}{8 \pi B} \sum_{i,j,k=0}^{2B-1} w_B(j) f(\alpha_i,\beta_j,\gamma_k) \overline{D(l,m,m';\alpha_i,\beta_j,\gamma_k)}, ~~~~ |m|,|m'| \leq l,
\end{equation} 
with the sampling angles $\alpha_i \coloneqq i \pi / B$, $\beta_j \coloneqq (2j+1) \pi / 4B$, $\gamma_k \coloneqq \alpha_k = k \pi / B$, and the quadrature weights
\begin{equation}\label{eq:weights}
w_B(j) \coloneqq \frac{2\pi \sin\beta_{j}}{B^{2}} \sum_{i=0}^{B-1} \frac{\sin((2i+1)\beta_{j})}{2i+1}.
\end{equation}
}
\indent
In the algorithmical context, any algorithm for computation of the Fourier coefficients $f^\circ(l,m,m')$ of functions $f \in H_B$ based on 
\eqref{eq:quadrature}
is referred to as a \emph{discrete \textnormal{SO(3)} Fourier transform}. 
Correspondingly, reconstruction of the function values $f(\alpha_i,\beta_j,\gamma_k)$ from the Fourier coefficients $f^\circ(l,m,m')$ via \eqref{eq:fourier_representation} is called an \emph{inverse} discrete SO(3) Fourier transform. 

\subsection{Fast SO(3) Fourier transform}
\label{subsec:sofft}

As can be seen with Gauss' well-known formula for the sum of the first $n$ positive integers, a function $f$ on SO(3) with bandwidth $B$ possesses $B(4B^2\!-1)/3$ potentially non-zero Fourier coefficients. One-by-one computation of these Fourier coefficients by evaluating the triple sum in \eqref{eq:quadrature} requires a total of $\mathcal{O}(B^6)$ computation steps. This is unacceptable for most practical purposes.

By a \emph{separation of variables}, \citet{Kostelec:Rockmore:2008} reduced this complexity: Inserting 
the expression \eqref{eq:Wigner_D} into \eqref{eq:quadrature} and rearranging the summands yields
\begin{align*}
f^\circ(l,m,m') = \frac{2l+1}{8 \pi B} \sum_{j = 0}^{2B-1} w_B(j) \hspace*{1pt} d(l,m,m';\beta_j) 
\underbrace{
\sum_{i,k=0}^{2B-1} f(\alpha_i,\beta_j,\gamma_k) \hspace*{1pt} \exp(\mathrm{i} (m \alpha_i + m' \gamma_k))
}_{\textstyle\hphantom{S(m,m';j)} \, \eqqcolon \, S(m,m';j)}\!.
\end{align*}
The following procedure lends itself well to this particular structure: In a first step, for all fixed $j = 0,\dots,2B-1$, the inner sum $S(m,m';j)$ is computed for all $m,m' = 1-B,\dots,B-1$ with a standard two-dimensional iFFT. This requires a total of $\mathcal{O}(B^3 \log B)$ steps. Then, for all fixed $m,m' = 1-B,\dots,B-1$, the Fourier coefficients $f^\circ(l,m,m')$, $l = \max\{|m|,|m'|\}, \dots, B - 1$, are computed. Due to the three-term recurrence relation \eqref{eq:Wigner_d_recurrence} of the Wigner-$d$ functions, this can be done in $\mathcal{O}(B^4)$ operations using, for example, the well-known Clenshaw algorithm \citep{clenshaw}, or in even only $\mathcal{O}(B^3 \log^2\! B)$ steps by using a fast discrete polynomial transform (FDPT, see \citep{potts_steidl_tasche,driscoll_healy_rockmore}). The symmetries \eqref{eq:Wigner_symmetries} can be used to speed up this second step. This, however, will not have an impact on the complexity. The complexity of the above-described sequential FSOFT thus amounts to $\mathcal{O}(B^4)$ or even only $\mathcal{O}(B^3 \log^2\! B)$.

In order to describe the corresponding fast inverse transform, we bring the second step above into matrix-vector notation. For this, let
\begin{align*}
V_B(m,m') &\coloneqq \mathrm{diag}\big[(2l+1)/8 \pi B\big]_{l\,=\,\max\{|m|,|m'|\},\dots,B-1}, \vphantom{\big[\big]_{\begin{subarray}{l}\{\}\\jB\end{subarray}}}\\ 
T_B(m,m') &\coloneqq \big[d(l,m,m';\beta_j)\big]_{\begin{subarray}{l}\hspace*{0.035cm}l\,=\,\max\{|m|,|m'|\},\dots,B-1\\[0.04cm]j\,=\,0,\dots,2B-1\end{subarray}},\\ 
W_B &\coloneqq \mathrm{diag}\big[w_B(j)\big]_{j\,=\,0,\dots,2B-1}. 
\end{align*}
Then, for all fixed $m,m'\! = 1-B,\dots,B-1$, it is
\begin{align*}
\big[f^\circ(l,m,m')\big]_{l\,=\,\max\{|m|,|m'|\},\dots,B-1} &= V_B(m,m') \, T_B(m,m') \, W_B \big[S(j; m, m')\big]_{j\,=\,0,\dots,2B-1},\\
\big[S(j; m, m')\big]_{j\,=\,0,\dots,2B-1} &= T_B(m,m')^\textnormal{T}\,  \big[f^\circ(l,m,m')\big]_{l\,=\,\max\{|m|,|m'|\},\dots,B-1}.
\end{align*}
Multiplication by the matrix $V_B(m,m')\,T_B(m,m')\,W_B$ is called the \emph{discrete Wigner transform} (DWT) of orders $m$ and $m'$ for the bandwidth $B$. Multiplication with the transposed matrix $T_B(m,m')^\textnormal{T}$ is thus referred to as the \emph{inverse} discrete Wigner transform (iDWT). It is easy to see how the iDWT and a two-dimensional FFT constitute the sequential iFSOFT, which has the same complexity as the FSOFT, when the iDWT is realized correspondingly with an adjoint Clenshaw algorithm or an adjoint FDPT.

\section{Parallelization}\label{sec:parallelization}
We use the well-known Forster method \citeyear{forster} to design our parallel FSOFT. The Forster method is also known as the PCAM method, which is an abbreviation for the four successive phases of \emph{partitioning}, \emph{communication}, \emph{agglomeration}, and \emph{mapping}, of which Forster's method is comprised. For more information on this method, we refer the reader to \citep{Quinn:2003,Roosta:2000,JaJa:1992}. Once the parallel FSOFT is designed, the corresponding parallel iFOSFT can be deduced from it directly. We shall not discuss parallelization of the two-dimensional FFT/iFFT, for this is not our concern (see Sects.\ \ref{sec:benchmark_tests} \& \ref{sec:discussion_and_conclusion}).

\emph{Partitioning.} As explained above, in the second step of the FSOFT, the DWT has to be performed for all different orders $m,m'\!=1-B,\dots,B-1$. This $m$-$m'$ partitioning is reasonably fine.

\emph{Communication.} As regards the underlying mathematics, a single DWT does not require any information of or from another DWT, and there is thus no such communication necessary between the threads. However, it clearly makes sense to exploit the seven Wigner-$d$ symmetries \eqref{eq:Wigner_symmetries} in order to speed up the computations. This results in communication between eight DWTs each, except for when $m=0$, $m'\!=0$, or $m = m'$. In these cases, the DWT groups are smaller, because not all seven Wigner-$d$ symmetries \eqref{eq:Wigner_symmetries} are meaningful here.

\emph{Agglomeration.} Drawing the required DWTs for a bandwidth $B$ into a two-dimensional $m$-$m'$ integer coordinate system reveals a quadratic area with side length $2B-1$. A direct approach for agglomeration would be to cluster the DWTs with respect to $m$ or $m'$ to groups of $2B-1$ each. However, as indicated above, pursuing such straightforward strategy would effectively prevent us from making use of the Wigner-$d$ symmetries \eqref{eq:Wigner_symmetries}. We thus agglomerate the DWTs based on these symmetries to groups of eight or less instead. No communication is required between these groups.

\emph{Mapping.} Since the distinct work packages of eight DWTs or less are relatively small, we assign them one-by-one to the available computation nodes, as explained in the following. 

Drawing all the DWT clusters into the same coordinate system as above results in a triangular area to be worked through; on the code level, this is represented by two nested `for' loops: $m = 0,\dots,B-1$ and $m'\! = 0,\dots,m$. Using Gauss' formula for the sum of the first $n$ positive integers, the two indices $m$ and $m'$ can be mapped bijectively onto a single linear index
\begin{equation}\label{eq:sigma}
\sigma \coloneqq \frac{m(m+1)}{2} + m'.
\end{equation}
This has the apparent advantage that the arising `for' loop over $\sigma = 0,\dots, B(B+1)/2$ can easily be worked through by the nodes. However, when reconstructing 
\begin{equation}\label{eq:m_mp_sigma}
\begin{aligned}
&m\hspace{-10pt} &&= m(\sigma)\hspace{-10pt} &&= \bigg\lfloor \sqrt{2\sigma + \frac{1}{4}} - \frac{1}{2} \bigg\rfloor,\\
&m'\! \hspace{-10pt}&&= m'(\sigma)\hspace{-10pt} &&= \sigma - \frac{m(\sigma)(m(\sigma)+1)}{2},
\end{aligned}
\end{equation}
floating-point arithmetic is required and square roots are to be taken.

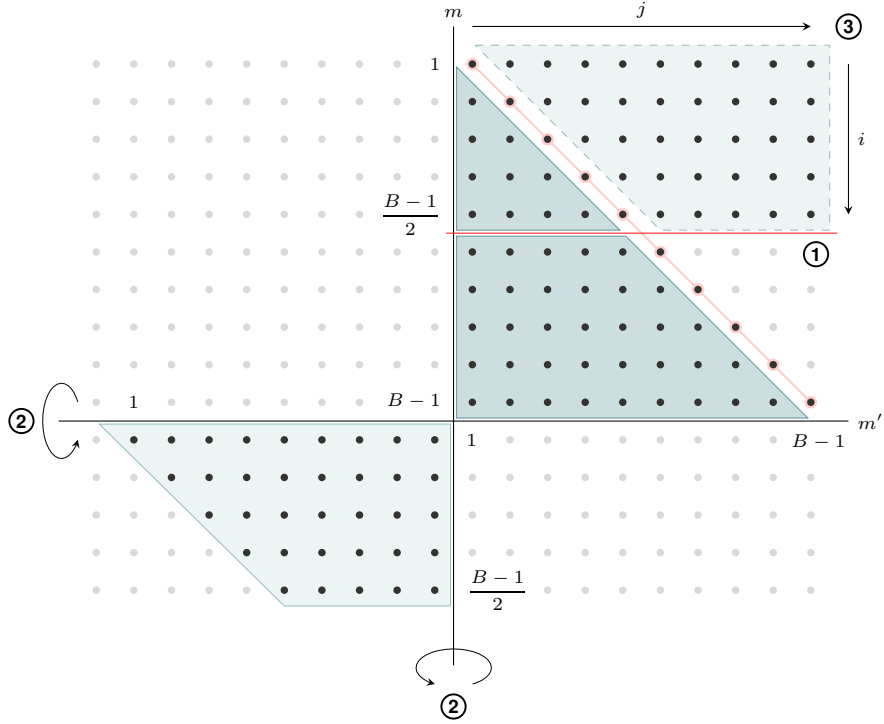
\begin{figure}[!t]
	\centering
	\begin{tikzpicture}[font=\sffamily\ssmall]
		\fill[solid, draw=UzLGreen!60, fill=UzLGreen!20] (5.79, 7.29) -- (5.79, 9.46) -- (7.96, 7.29) -- cycle;
		\fill[solid, draw=UzLGreen!60, fill=UzLGreen!20] (5.79, 4.79) -- (5.79, 7.21) -- (8.04,7.21) -- (10.46,4.79) -- cycle;	
		\fill[solid, draw=UzLGreen!40, fill=UzLGreen!7] (5.71, 4.71) -- (1.04, 4.71) -- (3.5, 2.29) -- (5.71, 2.29) -- cycle;
		\fill[dashed, draw=UzLGreen!40, fill=UzLGreen!7]  (6.04,9.75) -- (10.75,9.75) -- (10.75,7.29) -- (8.5, 7.29) -- cycle;
			
		\draw[thick, draw=red!20] (6, 9.5) to (10.5, 5);
		
		\foreach \x in {-9,...,10}
		{
           	\foreach \y in {6,...,20}
           	{
				\ifthenelse{\x<\y \AND \y>5 \AND \x<11 \AND \y<11 \AND \x>0}
					{\node[dot] (\x-\y) at (-0.5 * \x + 6, 0.5 * \y - 0.5){};}
					{\node[brightdot] (\x-\y) at (-0.5 * \x + 6, 0.5 * \y - 0.5){};}
			}
		}
		
		\foreach \x in {1,...,10}
		{
           	\foreach \y in {1,...,10}
           	{
				\ifnum \x=\y
					\node[bigger_brighter_dot] (\x-\y) at (5 - 0.5 * \x + 6, 0.5 * \y + 4.5){};
				\fi
			}
		}
			
		\foreach \x in {1,...,10}
		{
           	\foreach \y in {1,...,10}
           	{
				\ifthenelse{\x>\y \AND \y>5}
					{\node[brightdot] (\x-\y) at (0.5 * \x + 5.5, -0.5 * \y + 10){};}
					{\node[dot] (\x-\y) at (0.5 * \x + 5.5, -0.5 * \y + 10){};}
       		}
		}
		
		\node at (5.75,1.5) {\AxisRotator[rotate=90]};
		\node at (0.5,4.75) {\AxisRotator[rotate=180]};
		
		\draw[solid, black](5.75,1.5) -- (5.75,10);
		\draw[solid, black](0.5,4.75) -- (11,4.75);
		
		\draw[solid, red!90](5.65, 7.25) -- (10.85, 7.25);
		
		\node[left=0.3cm, fill=white] at (6,9.5) {$1$};
		\node[above left=0.3cm and 0.3cm, fill=white] at (6,4.5) {$B-1$};
		\node[below right=0.3cm and 0.3cm, fill=white] at (5.5,5) {$1$};
		\node[above=0.3cm, fill=white] at (1.5,4.5) {$1$};
		\node[below=0.3cm, fill=white] at (10.6,5) {$B-1$};	
		\node[above] at (5.75,10) {$m$};	
		\node[below] at (5.75,1.25) {\circled{2}};	
		\node[right] at (11,4.75) {$m'$};
		\node[left] at (0.3,4.75) {\circled{2}};
		\node[below left=0.1cm and 0.1cm, fill=white, inner sep=0pt] at (10.85, 7.25) {\circled{1}};
		\node[left, fill=white, inner sep=.2em] at (5.65, 7.5)  {\color{black}$\displaystyle\frac{B-1}{2}$};
		\node[right, fill=white, inner sep=.2em] at (5.86, 2.5){\color{black}$\displaystyle\frac{B-1}{2}$};
		
		\draw[-stealth] (11,9.5) -- (11,7.5) node [midway, right] {$i$};
		\draw[-stealth] (6,10) -- (10.5,10) node [midway, above] {$j$};
		\node[] at (11,10) {\circled{3}};
		
	\end{tikzpicture}
	\begin{minipage}{0.95\textwidth}
	\centering
	\caption{Geometric approach for optimization of the nested `for' loops over $m=0,\dots,B-1$ and $m'\!=0,\dots,m$. The triangular area shaded in darker green contains the groups of DWTs for all combinations of $m$ and $m'$ for a given odd bandwidth $B$, except for those groups with $m=0$, $m'\!=0$, or $m=m'$. The DWTs are clustered according to the Wigner-$d$ symmetries \eqref{eq:Wigner_symmetries}. \circled{\textsf{1}} The triangular area is divided into two parts by a cut at halfway along the $m$ axis. \circled{\textsf{2}} The lower part of the triangular area is mirrored at both axes. \circled{\textsf{3}} The mirrored area fits in the empty space in the upper half of the square with side length $B-1$ containing the original triangle. The resulting rectangle can be worked through using two independent indices $i$ and $j$. In doing so, the groups with $m = m'$ are skipped. We treat these cases separately because not all seven Wigner-$d$ symmetries \eqref{eq:Wigner_symmetries} are meaningful for these groups. The same applies to the groups with $m=0$ or $m'\!=0$, which we also treat in advance.}\label{fig:par_opt}
	\end{minipage}
\end{figure}

By a purely geometric approach, we can transform the above-described triangular index range bijectively into a rectangular area in such a way that reconstruction of the indices $m$ and $m'$ is less complicated. Our idea is illustrated in Fig.\,\ref{fig:par_opt}. Here, we assume that the bandwidth $B$ is odd; however, the following formulae are also valid for an even band\-width. 
The indices $m=1,\dots,B-1$ and $m'\!=1,\dots,m-1$ can be reconstructed from the indices $i = 1,\dots,\lfloor(B-1)/2\rfloor$ and $j = 1,\dots,B-1$ introduced in Fig.\,\ref{fig:par_opt} via
\begin{alignat*}{3}
&m &&= m(i,j) &&= 
\begin{cases}
B - i &\textnormal{if }\hspace*{1pt} j > i,\\
i + 1 &\textnormal{otherwise},
\end{cases}\\
&m'\! &&= m'(i,j) &&=
\begin{cases}
B - j &\textnormal{if }\hspace*{1pt} j > i,\\
j     &\textnormal{otherwise}
\end{cases}
\end{alignat*}
(for $B$ odd and $i = (B-1)/2$ only the values $j=1,\dots,(B-1)/2$ are actually needed).
Note that only a conditional `if' statement and integer arithmetics are required here, as opposed to in \eqref{eq:m_mp_sigma}.
The indices $i$ and $j$ can now be mapped bijectively onto a single linear index (cf.\ \eqref{eq:sigma}) 
\begin{equation*}
\kappa \coloneqq (i-1)(B-1) + (j - 1),
\end{equation*}
so that they can be reconstructed via
\begin{alignat*}{2}
i &= i(\kappa) &&= \left\lfloor \frac{\kappa}{B-1} \right\rfloor + 1,\\
j &= j(\kappa) &&= \textnormal{mod}(\kappa,B - 1) + 1.
\end{alignat*}
Here, only integer division as well as modulus and increment operations are required.
The resulting `for' loop over $\kappa = 0,\dots,(B-1)(B-2)/2$ can now be parallelized, so that the DWT clusters are distributed efficiently to the available computation nodes.

The thus finished parallel FSOFT scales well with the problem size and the number of available computation nodes. The reason for this lies within the small size of the working packages and the fact that there is no communication required between the working threads. This in turn yields a high locality of the data. Memory access of the different nodes can be made exclusive, in the sense that each node works in its own memory range. 
Furthermore, the fine granularity of the working packages makes it easy for the compiler to find a good scheduling. As indicated above, it is now clear how to derive the corresponding parallel iFSOFT. With some additional effort, it can be shown that the above-described parallel FSOFT and its inverse are \emph{work-optimal} (cf.\ \citep[P.\,32]{JaJa:1992})\,--\,see \citep[Sec.\,2.5]{bachelor_denis} for proof.

\section{Benchmark test}\label{sec:benchmark_tests}
\input{numerical_experiments}

\section{Discussion}\label{sec:discussion_and_conclusion}
The results of the benchmark test clearly show that our parallelization of the FSOFT and iFSOFT was successful, and that current and future applications can benefit considerably from these novel parallel algorithms and our corresponding 
C++ implementation using OpenMP.

As can be seen by having a closer look at the speedup depicted in Fig.\,\ref{fig:speedup}, the beneficial impact of our parallelization initially increases fast with the bandwidth, especially when many nodes are used. This effect is stronger in the forward than in the inverse transform. The results 
suggest that there is a limiting bandwidth to this effect, however\,--\,in both the forward and inverse transform, the bandwidth 512 already presents a slightly lower speedup than the bandwidth 256; this can be attributed to increased side effects (see below). The runtime of the iFSOFT was generally longer than that of the FSOFT (see Fig.\,\ref{fig:runtimes}). The reason for this is the matrix transposition in the current version of the iDWT. This transposition of matrices of strongly varying size is performed on-the-fly by the computation nodes, which especially in the larger bandwidths results in a lower speedup in the inverse transform due to common side effects such as increasing workload imbalance and memory management overhead. The time for computing the SO(3) quadrature weights \eqref{eq:weights} in the forward transform, on the other hand, is negligible short. Generally, in all bandwidths considered, the speedup initially increases fast with the number of computation nodes, and in the larger bandwidths 128, 256, and 512 even almost reaches its optimum (equal to the number of computation nodes), until around eight nodes. Subsequently, the speedup begins to plateau, again due to influencing factors such as increasingly complicated memory management, etc. This is also reflected in the naturally decreasing efficiency shown in Fig.\,\ref{fig:efficiency}. While it is conceivable that our parallel algorithms will benefit from employing more nodes than available on our testing system, the results overall suggest that it would be of no avail to recruit a very much higher number of computation nodes with a comparable system architecture, at least until a faster DWT/iDWT is used.

As indicated above, an elaborate parallelization of the two-dimensional FFT/iFFT is beyond the scope of this paper. Works on parallelization of the classical FFTs include \citep{pfft,pfft3,pfft2}. Although the speedup achieved in these works cannot directly be compared with that in our case, it is interesting to see the very similar behavior in \citep[Fig.\,5]{pfft} (note that the number of relevant SO(3) Fourier coefficients for the bandwidth $B$ is $B(4B^2-1)/3$, while the number of Fourier coefficients in a classical three-dimensional FFT with bandwidth $N$ is $N^3$. Moreover, the cost is different for computing a single Fourier coefficient because the 
more demanding DWT replaces a classical DFT, simply put). It is noteworthy that particularly in the larger bandwidths, 
the runtime of the parallel two-dimensional iFFT/FFT proposed by the FFTW developers constituted only small portion of the total runtime of the FSOFT/iFSOFT in our benchmark test (approximately 5\% and 8\%, respectively, for the bandwidth 512 when using all 64 
cores). 

Apart from our parallelization, an additional contribution of this paper to computational harmonic analysis on SO(3) in general is our treatment of the both accuracy- and memory-critical bandwidth 512. To our knowledge, this large bandwidth has not been attempted by others before (compare especially \citep{mcewen_et_al,Prestin:2009,Kostelec:Rockmore:2008}). The results of the error measurement in Table \ref{tab:error} show that we were successful in performing the full forward and inverse transform for this large bandwidth. This is due to the large amount of RAM available on our testing system and, more importantly, the fact that we increased from double to extended double precision (double precision is not sufficient). The errors shown in Table \ref{tab:error} for the other bandwidths are much smaller than those in \citep[Table 7]{Kostelec:Rockmore:2008}; they are not directly comparable with those in \citep[Fig.\,6]{Prestin:2009}, because the error was defined differently by these authors. As mentioned above, the bandwidth 512 also benefits greatly from our parallelization, resulting in a speedup of 34.34 (forward) and 24.25 (inverse) using all 64 available cores, resulting from a runtime of approximately three minutes (forward) and 4.3 minutes (inverse) as opposed to 1.53 hours (sequential forward) and 1.74 hours (sequential inverse).

Now that we realized our idea and demonstrated its feasibility as a general proof of concept already showing practical applicability, the next version of our 
software will include a faster DWT/iDWT based on Clenshaw's algorithm \citeyear{clenshaw}.


\bibliographystyle{ACM-Reference-Format-Journals}
\footnotesize\bibliography{references}

\received{February 20XX}{March 20XX}{June 20XX}

\end{document}